\documentclass[a4paper]{jpconf}
\usepackage{amsmath}
\usepackage{amssymb}
\usepackage{pifont}
\usepackage{graphicx}
\begin{document}
\title{Top couplings and new physics: theoretical overview and developments}

\author{Cen Zhang}

\address{
Department of Physics, Brookhaven National Laboratory, Upton, N.Y., 11973, U.S.A.
}

\ead{cenzhang@bnl.gov}

\begin{abstract}
  Top-quark physics has entered the precision era.  In this talk we discuss the
  theoretical ingredients required for a global approach to the complete set of
  top-quark couplings at NLO accuracy.  In particular, recent developments on
  top-quark flavor-changing neutral couplings are shown as an example.
  Aspects of flavor-conserving sector will also be discussed.
\end{abstract}

\section{Introduction}

The top quark and the Higgs boson are the heaviest known fundamental particles.
These two particles and their mutual interactions can play crucial roles in
many extensions of the standard model (SM).  While global analyses of the Higgs
interactions based on the effective field theory (EFT) approach are available
in literatures, the top quark couplings have been rarely studied with
the same strategy.  In this talk we will discuss the theoretical framework for
a global approach to top quark couplings, in particular with a focus on
higher-order QCD corrections and their impacts on the global strategy.

\section{Towards a global fit at NLO}

The millions of top quarks produced at the Tevatron and the LHC have brought top
physics to its precision era.  In the past year, progresses have been made in
the measurements of various top processes, thanks to the 8 TeV data set at the LHC.
The single top production in $Wt$ channel has been observed for the first time
\cite{Chatrchyan:2014tua}.  Several measurements and searches on the $t\bar tX$
final state have been updated
\cite{Khachatryan:2014ewa,CMS:2014wma,Khachatryan:2014qaa,Aad:2014lma}, and new
constraints on the flavor-changing neutral couplings, $tq\gamma$ and $tqh$,
become available \cite{CMS:2014hwa,CMS:2014qxa,Aad:2014dya}.

To extract top-quark couplings from these measurements, the EFT provides a
suitable theoretical framework.  While EFT has been widely used as a
model-independent description of new physics, here we want to emphasize that
this approach has the advantage of being consistent at higher orders in
perturbative calculation.  This is an important feature for top physics, as
QCD corrections are in general not negligible for top-quark processes at hadron
colliders, in particular when non-SM couplings are involved.  Furthermore, in
certain processes these couplings are accessible only through a top-quark loop.
Therefore a reliable analysis requires predictions within the EFT framework at
NLO accuracy.

In an EFT approach, deviations from the SM are described by adding
higher-dimensional gauge-invariant operators to the SM Lagrangian.  These
operators are suppressed by inverse powers of $\Lambda$, where $\Lambda$ is the
scale at which new physics lives.  The theory can be renormalized order by order
in $1/\Lambda$, provided that all operators up to a certain order are incorporated.
Thus perturbative calculations can be systematically improved, to any desired order
of $(\alpha_s/\pi)^m(1/\Lambda)^n$.

It is important to keep in mind that the EFT is a consistent approach only if
the complete set of operators are taken into account. In particular, turning on
one or several operators at a time may lead to misleading or even
basis-dependent conclusions.  One example is the two ``blind directions''
appeared in precision electroweak fit \cite{Grojean:2006nn}.  These directions
are transparent only in certain operator basis and can be overlooked if one
turns on only one operator at a time.  This is one of the reasons why a global
analysis can be important.

\begin{figure}[t]
  \begin{minipage}{.45\linewidth}
    \begin{center}
      \includegraphics[width=.9\linewidth]{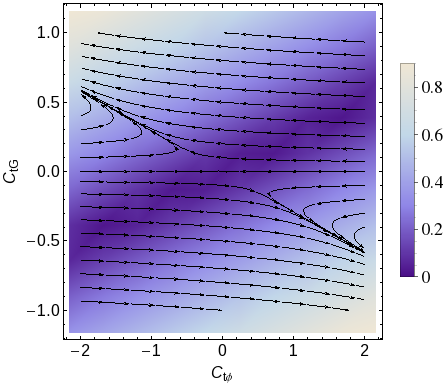}
    \end{center}
    \caption{\label{fig:rg} RG evolution of $O_{t\varphi}$ and $O_{tG}$.
  The arrows represent the direction of RG flows when the scale increases.  The color
  shows the shift in $C_{t\varphi}-C_{tG}$ space when evolved from $m_t$ to 2 TeV.}
  \end{minipage}
  \hfill
  \begin{minipage}{.5\linewidth}
    \begin{center}
      \includegraphics[width=.9\linewidth]{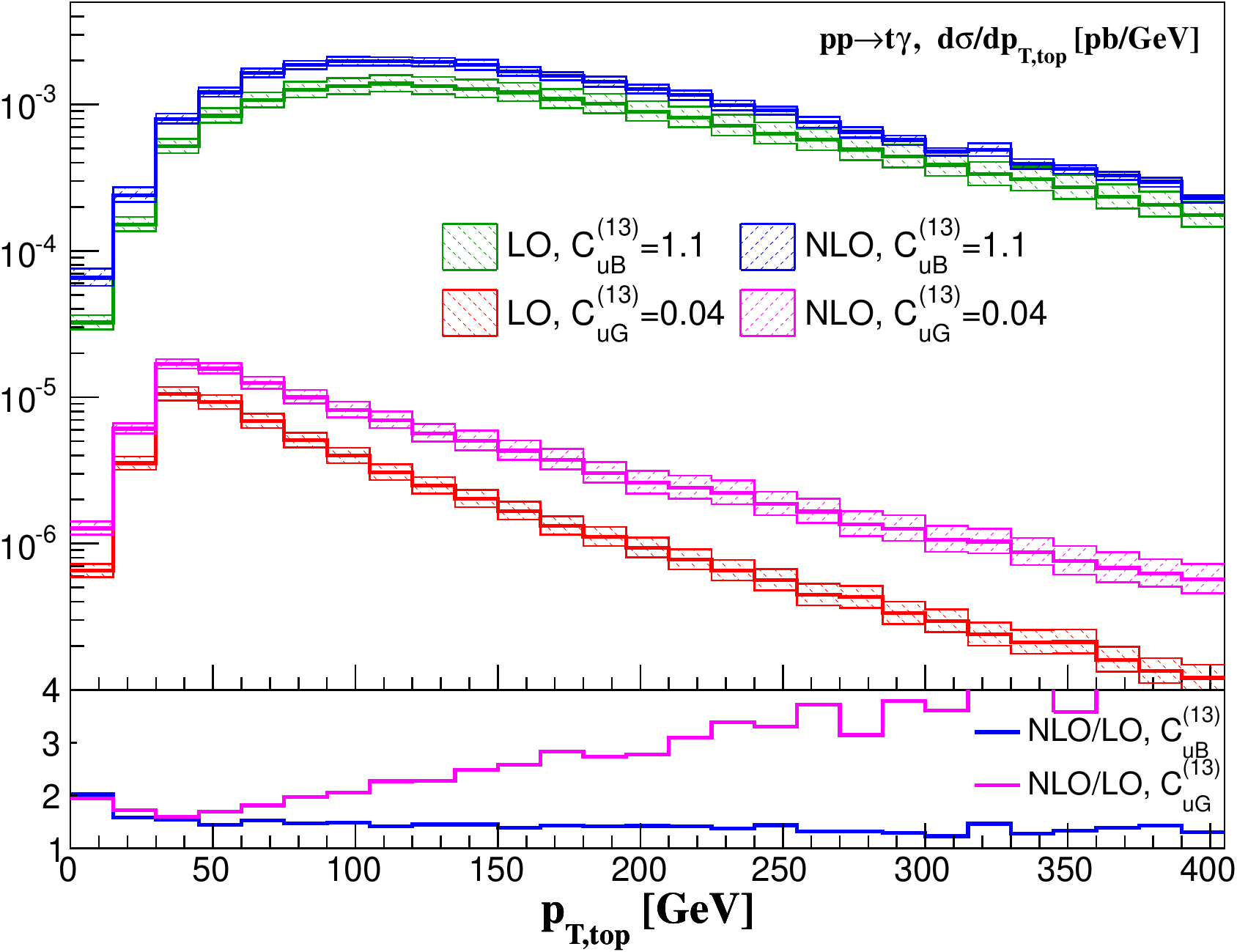}
    \end{center}
    \caption{\label{fig:pt} The $p_T$ distribution of top quark in $pp\to t\gamma$
    at $\sqrt{s}=8$ TeV.  Photon $p_T>30$ GeV and $\eta<2.5$ cuts are imposed.
    $\Lambda=1$ TeV. Coefficients are chosen within the current limits.}
  \end{minipage}
\end{figure}

At NLO, a global strategy becomes even more crucial, as the renormalization
group (RG) mixing effects clearly reveals the intrinsic relations among
different operators.  The evolution of operators are described by a matrix,
and mixing effects among different operator coefficients occur whenever the
renormalization scale changes.  Conceptually, this implies that the very
definition of top couplings are obscured at NLO, and the distinction between
individual couplings can depend on the scale at which they are probed.  This is
illustrated in Figure \ref{fig:rg}, where the RG flows of the operators
$O_{t\varphi}$ and $O_{tG}$ are displayed.  These two operators give rise to
anomalous $t\bar th$ and $t\bar tg$ couplings, respectively.

The top EFT has been studied at leading order (LO) for a long time (see, for
example, Ref.~\cite{Zhang:2010dr} for more details).  At NLO,
however, there are still many aspects that require further work.  QCD
corrections to most dimension-six operators remain to be studied, and for
certain processes these include higher-order contaminations from operators not
involved at the tree level.  In addition, simulation tools that are capable of
doing such calculations are desirable in order to have realistic analyses.  In
the next section we will give a brief summary of some recent progresses made in
the flavor-changing neutral current (FCNC) sector of the top quark.

\section{Flavor-changing sector}

Processes triggered by the FCNC couplings of the top quark are highly suppressed
in the SM by the GIM mechanism, thus any signal observed indicates new physics.
A wide variety of limits have been set on top-quark FCNC interactions.
Currently the best ones on trilinear couplings are from decay processes $t\to ql^+l^-$,
$t\to qh$, and production processes $qg\to t$ and $qg\to t\gamma$, all have
been searched at the LHC.  On the other hand, single top production at LEP2 could 
provide limits on contact four-fermion interactions.

In an EFT the FCNC couplings involving one top-quark and one light-quark field
arise from the following dimension-six operators:
\newcommand{\FDF}{\left(\varphi^\dagger\overleftrightarrow{D}_\mu\varphi\right)}
\newcommand{\FDFI}{\left(\varphi^\dagger\overleftrightarrow{D}^I_\mu\varphi\right)}
\vspace{-6pt}
\begin{equation}
  \begin{array}{lll}
    O_{\varphi q}^{(3,i+3)}=i\FDFI(\bar{q}_i\gamma^\mu\tau^IQ)
    &
    O_{\varphi q}^{(1,i+3)}=i\FDF(\bar{q}_i\gamma^\mu Q) 
    &
    O_{\varphi u}^{(i+3)}=i\FDF(\bar{u}_i\gamma^\mu t)
    \\
    O_{uB}^{(i3)}=g_Y(\bar{q}_i\sigma^{\mu\nu}t)\tilde{\varphi}B_{\mu\nu}
    &
    O_{uW}^{(i3)}=g_W(\bar{q}_i\sigma^{\mu\nu}\tau^It)\tilde{\varphi}W^I_{\mu\nu}
    &
    O_{uG}^{(i3)}=g_s(\bar{q}_i\sigma^{\mu\nu}T^At)\tilde{\varphi}G^A_{\mu\nu}
    \\
    O_{u\varphi}^{(i3)}=(\varphi^\dagger\varphi)(\bar{q}_it)\tilde\varphi
    \nonumber
  \end{array}
  \label{eq:allfcnc}
\vspace{-6pt}
\end{equation}
where the subscript $i=1,2$ is the generation of the quark field. $Q$ is the
third-generation doublet. For operators with $(i3)$ superscript, a similar set
of operators with $(3i)$ flavor structure can be obtained by interchanging
$(i3)\leftrightarrow (3i),\ t\leftrightarrow u_i$ and $Q\leftrightarrow q_i$.
We further define $O_{\varphi q}^{(-,i+3)}$ as $(O_{\varphi
q}^{(1,1+3)}\!-\!O_{\varphi q}^{(3,1+3)})/2$.  In addition, four-fermion
operators including two quarks and two leptons are also relevant.  A full list
can be found in Ref.~\cite{Zhang:2014rja}.

It is worth pointing out that the limits obtained by experimental collaborations
almost always assume one single FCNC interaction is present at a time.  In
addition, four-fermion operators are neglected in most cases.  These operators
could for example describe models where FCNC couplings are mediated by new heavy
particles.  While they could obviously contribute to $e^+e^-\to tj$, their
effects in top decay $t\to qZ\to ql^+l^-$ are not negligible either, even
after applying the $Z$-mass window cuts on the lepton pairs.  To have a complete
understanding of the current status, one should follow a global approach
where all FCNC operators are turned on simultaneously. 

Recently, theoretical ingredients required for such a global analysis have
been completed at NLO accuracy.  In the FCNC sector the relevant mixing effects
are those among $O_{uB}^{(i3)}$, $O_{uW}^{(i3)}$, $O_{uG}^{(i3)}$ and
$O_{u\varphi}^{(i3)}$, and their $(3i)$ counterparts.  NLO predictions for FCNC
processes are now available with the operator mixing effects properly taken
into account.  Various of decay processes have been computed in
Refs.~\cite{Zhang:2014rja}.  In these processes the $\mathcal{O}(\alpha_s)$
corrections come from not only the standard QCD corrections to dimension-six
operators, but also from the operators $O_{uG}^{(i3)}$ and $O_{uG}^{(3i)}$,
which affect top decays only at NLO.  Contributions from four-fermion operators
in three-body decays are also included.  On the production side, single top
production $qg\to t$ has been computed in Ref.~\cite{Liu:2005dp}.  More
recently, the full set of two-fermion FCNC operators, listed in
Eq.~(\ref{eq:allfcnc}), have been implemented in the MG5\_aMC@NLO framework
\cite{Alwall:2014hca}, using FeynRules \cite{Alloul:2013bka} and NLOCT
\cite{Degrande:2014vpa}.  Details can be found in Ref.~\cite{aMCfcnc}.  This
allows for processes such as $pp\to tX$, with $X=\gamma, Z, h$, to be computed
automatically at NLO and matched to parton shower simulation.  For illustration
we show in Figure~\ref{fig:pt} the $p_T$ distribution in $pp\to t\gamma$ at 8
TeV center of mass energy.  The $e^+e^-\to tj$ process can be studied in the
same framework, with the caveat that the four-fermion operators are no yet
available.  They are planned to be implemented in future, and for the moment
analytical results for these operators can be used.

\begin{figure}[t]
  \begin{minipage}{.49\linewidth}
    \begin{center}
      \includegraphics[width=.8\linewidth]{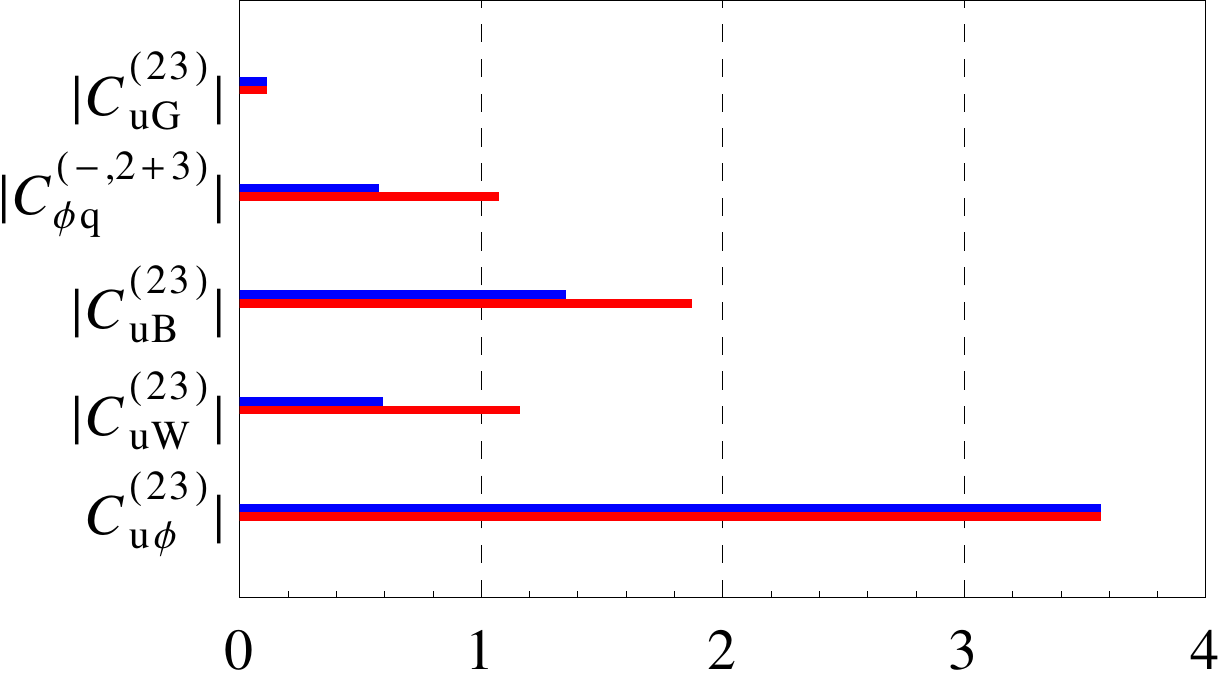}
    \end{center}
  \end{minipage}
  \hfill
  \begin{minipage}{.49\linewidth}
    \begin{center}
      \includegraphics[width=.8\linewidth]{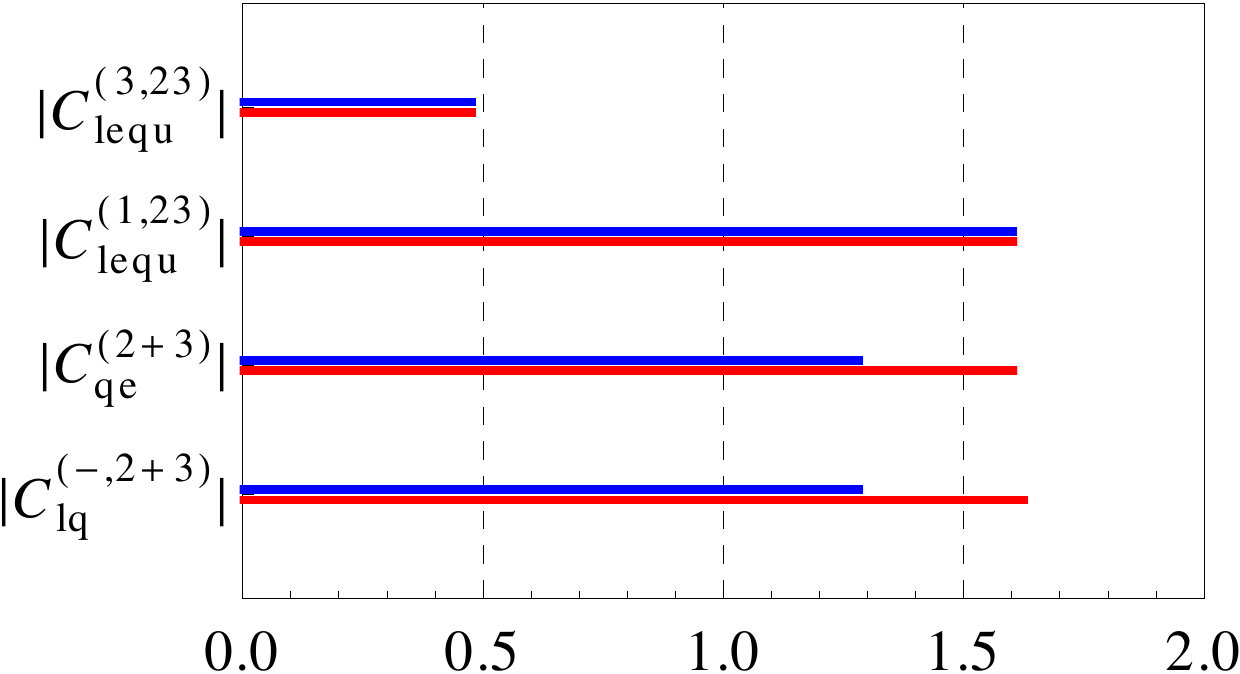}
    \end{center}
  \end{minipage}
    \caption{\label{fig:limits}Left: limits on two-fermion operator
    coefficients. Blue lines indicate limits obtained by setting other
    coefficients to zero. Red lines are obtained by allowing other coefficients
    to float. $\Lambda=1$ TeV is assumed. Right: limits on some four-fermion
    operators.  Their definitions can be found in Ref.~\cite{fcncfit}.}
\end{figure}

With the above results, a global fit including all processes
mentioned in this section can be performed.  From the published information it
is not clear to us how to combine the 95\% CL bounds from different measurements
consistently.  However, to illustrate the feasibility of this approach, a toy
fit is straightforward, by naively combining available limits on branching
ratios and cross sections.  In Figure \ref{fig:limits} we show some results
from the toy fit. The blue lines are obtained by setting other coefficients to zero,
while the red lines are obtained by allowing other coefficients to float. For
some operators the blue and red lines are different, indicating that a
correlation is present between the operators. These information are available
from the global fit. The complete analysis is presented in
Ref.~\cite{fcncfit}.

\section{Flavor-conserving sector}

The non-FCNC sector is more complicated.  The top couplings are parameterized by the
following operators \cite{AguilarSaavedra:2008zc,AguilarSaavedra:2009mx}
\vspace{-6pt}
\begin{equation}
  \begin{array}{lll}
    O_{\varphi Q}^{(3)}=i\FDFI(\bar{Q}_i\gamma^\mu\tau^IQ)
    &
    O_{\varphi Q}^{(1)}=i\FDF(\bar{Q}_i\gamma^\mu Q) 
    &
    O_{\varphi t}=i\FDF(\bar{t}_i\gamma^\mu t)
    \\
    O_{\varphi\varphi}=i\FDF(\bar{t}_i\gamma^\mu b)
    &
    O_{tB}=g_Y(\bar{Q}_i\sigma^{\mu\nu}t)\tilde{\varphi}B_{\mu\nu}
    &
    O_{tW}=g_W(\bar{Q}_i\sigma^{\mu\nu}\tau^It)\tilde{\varphi}W^I_{\mu\nu}
    \\
    O_{bW}=g_W(\bar{Q}_i\sigma^{\mu\nu}\tau^Ib){\varphi}W^I_{\mu\nu}
    &
    O_{tG}=g_s(\bar{Q}_i\sigma^{\mu\nu}T^At)\tilde{\varphi}G^A_{\mu\nu}
    &
    O_{t\varphi}=(\varphi^\dagger\varphi)(\bar{q}_it)\tilde\varphi
    \nonumber
  \end{array}
  \label{eq:allfc}
\vspace{-6pt}
\end{equation}
Four-fermion operators are not listed here.

\begin{table}
  \newcommand{\lmark}{L}
  \newcommand{\nmark}{N}
  \caption{Top-quark operators and some of the key processes at the LHC. 
    $O_{4f}$ denotes any four-fermion operator. 
  ``L'' (``N'') represents LO (NLO) contribution.
  $O_{G}$ and $O_{\varphi G}$ are included. $O_{\varphi\varphi}$ and
  $O_{bW}$ are omitted as their leading contributions are suppressed by $b$
  mass.}
  \begin{center}
    \begin{tabular}{lcccccccccc}
      \br
      Process & $O_{tG}$ & $O_{tB}$ & $O_{tW}$ & $O_{\varphi Q}^{(3)}$ &
      $O_{\varphi Q}^{(1)}$ & $O_{\varphi t}$ & $O_{t\varphi}$ &
      $O_{\mbox{4f}}$ & $O_{G}$ & $O_{\varphi G}$
      \\\mr
      $t\to bW\to bl^+\nu$
      &\nmark& &\lmark & \lmark &&&& \lmark && \\
      $pp\to t\tilde q$ &\nmark & &\lmark & \lmark & & & &\lmark & &
      \\
      $pp\to tW$
      &\lmark & &\lmark & \lmark &&&&\nmark&\nmark&\nmark
      \\
      $pp\to t\bar t$ 
      &\lmark&&&&&&\nmark&\lmark&\lmark&\lmark
      \\
      $pp\to t\bar t\gamma$
      &\lmark &\lmark &\lmark & & & &\nmark&\lmark&\lmark&\lmark
      \\
      $pp\to t\bar tZ$
      &\lmark &\lmark &\lmark &\lmark &\lmark &\lmark &\nmark&\lmark&\lmark&\lmark
      \\
      $pp\to t\bar th$
      &\lmark & & &&& &\lmark&\lmark&\lmark&\lmark
      \\
      {$gg\to H$, $H\to \gamma\gamma$}
      &\nmark & & &&& &\nmark&&&\lmark
      \\\br
    \end{tabular}
  \end{center}
\end{table}

Table 1 displays some of the key processes at the LHC, together with the
relevant operators in each process.  Most two-fermion operators are constrained
to some level, thanks to the recent LHC data.  However, more aspects need to be
studied before a real global fit can be made.  In a global approach a complete
operator basis should be used, which means that one should include not only
four-fermion operators that are often neglected, but also operators with no
top-quark field, such as
\begin{equation}
  O_G=g_sf^{ABC}G_{\mu}^{A\nu}G_{\nu}^{B\rho}G_{\rho}^{C\mu},
  \qquad
  O_{\varphi G}=g_s^2\left( \varphi^\dagger\varphi \right)
  G^A_{\mu\nu}G^{A\mu\nu}
  \ .
\end{equation}
They are included in Table 1, and could indeed affect most top processes.

Mixing effects may be important.  For instance, a non-vanishing operator
$O_{tG}$ at 1 TeV scale can induce an operator $O_{t\varphi}$ at the scale
$m_t$, with $C_{t\varphi}(m_t)\approx0.45C_{tG}(1\ \mathrm{TeV})$ from the QCD
mixing term.  Moreover, operators without a top-quark field, such as $O_G$ and
$O_{\varphi G}$, can be relevant through mixing effects.  In particular the
anomalous dimensions $\gamma_{tG,\varphi G}$, $\gamma_{tG,G}$, and
$\gamma_{\varphi G,tG}$ are non zero (the last one has been studied in
Ref.~\cite{Degrande:2012gr}).

Once the mixing is understood, complete NLO calculation can be carried out.
While some operators only give rise to SM-like couplings,
others are essentially of higher dimension, and QCD corrections are unknown in
most cases, with a few exceptions such as top-decay processes as well as
four-fermion operators in $t\bar{t}$ production.  From the study of FCNC sector
we see that these corrections can be potentially large, and so it is important
know their sizes.  The feasibility of implementing these operators in the
MG5\_aMC@NLO framework is being investigated.

\section{Summary}

With a variety of top processes investigated at the LHC, global analyses based
on top EFT are expected to provide a complete understanding of top-quark
couplings. However, QCD corrections can be potentially important, and a
reliable analysis at NLO accuracy requires further theoretical efforts.  We
have shown that recent progress completes the theoretical predictions in the
top FCNC sector.  While the flavor-conserving sector can be more complicated,
the feasibility of the same approach is being studied, and automatic NLO
calculations involving higher-dimensional top-quark operators can be expected
in future.

\ack
I would like to thank my collaborators, Celine Degrande, Gauthier Durieux, Fabio
Maltoni and Jian Wang.  This work is supported in part by the IISN ``Fundamental
interactions'' convention 4.4517.08, and by US Department of Energy under Grant
DE-AC02-98CH10886.


\section*{References}
\begin{thebibliography}{9}

\bibitem{Chatrchyan:2014tua} 
  CMS Collaboration 2014
  %``Observation of the associated production of a single top quark and a W boson in pp collisions at sqrt(s) = 8 TeV,''
  {\it Phys. Rev. Lett.}  {\bf 112} 231802
  %%CITATION = ARXIV:1401.2942;%%
  %29 citations counted in INSPIRE as of 14 Nov 2014

\bibitem{Khachatryan:2014ewa} 
  CMS Collaboration 2014
  %``Measurement of top quark-antiquark pair production in association with a W or Z boson in pp collisions at $\sqrt{s} = 8$ $\,\text {TeV}$,''
  {\it Eur. Phys. J.} C {\bf 74} 3060
  %%CITATION = ARXIV:1406.7830;%%
  %3 citations counted in INSPIRE as of 14 Nov 2014

\bibitem{CMS:2014wma} 
  CMS Collaboration 2014
  Measurement of the inclusive top-quark pair + photon production cross section in the muon + jets channel in pp collisions at 8 TeV
  {\it CMS-PAS-TOP-13-011}
  %%CITATION = CMS-PAS-TOP-13-011;%%
  %4 citations counted in INSPIRE as of 14 Nov 2014

\bibitem{Khachatryan:2014qaa} 
  CMS Collaboration 2014
  %``Search for the associated production of the Higgs boson with a top-quark pair,''
  {\it JHEP} {\bf 1409} 087
  [Erratum-ibid.\  {\bf 1410} 106]
  %%CITATION = ARXIV:1408.1682;%%
  %5 citations counted in INSPIRE as of 14 Nov 2014

\bibitem{Aad:2014lma} 
  ATLAS Collaboration 2014
  Search for $H \to \gamma\gamma$ produced in association with top quarks and constraints on the Yukawa coupling between the top quark and the Higgs boson using data taken at 7 TeV and 8 TeV with the ATLAS detector
  {\it Preprint 1409.3122 [hep-ex]}
  %%CITATION = ARXIV:1409.3122;%%
  %2 citations counted in INSPIRE as of 14 Nov 2014

\bibitem{CMS:2014hwa} 
  CMS Collaboration 2014
  Search for anomalous single top quark production in association with a photon
  {\it CMS-PAS-TOP-14-003}
  %%CITATION = CMS-PAS-TOP-14-003;%%
  %6 citations counted in INSPIRE as of 21 Nov 2014

\bibitem{CMS:2014qxa} 
  CMS Collaboration 2014
  Combined multilepton and diphoton limit on t to cH
  {\it CMS-PAS-HIG-13-034}
  %%CITATION = CMS-PAS-HIG-13-034;%%
  %16 citations counted in INSPIRE as of 21 Nov 2014

\bibitem{Aad:2014dya} 
  ATLAS Collaboration 2014
  %``Search for top quark decays $t \to qH$ with $H \to \gamma\gamma$ using the ATLAS detector,''
  {\it JHEP} {\bf 1406} 008
  %[arXiv:1403.6293 [hep-ex]].
  %%CITATION = ARXIV:1403.6293;%%
  %16 citations counted in INSPIRE as of 21 Nov 2014

\bibitem{Grojean:2006nn} 
  Grojean C, Skiba W and Terning J 2006
  %``Disguising the oblique parameters,''
  {\it Phys. Rev.} D {\bf 73} 075008
  %[hep-ph/0602154].
  %%CITATION = HEP-PH/0602154;%%
  %39 citations counted in INSPIRE as of 22 Nov 2014

\bibitem{Zhang:2010dr} 
  Zhang C and Willenbrock S 2011
  %``Effective-Field-Theory Approach to Top-Quark Production and Decay,''
  {\it Phys. Rev.} D {\bf 83} 034006
  %[arXiv:1008.3869 [hep-ph]].
  %%CITATION = ARXIV:1008.3869;%%
  %74 citations counted in INSPIRE as of 22 Nov 2014

\bibitem{Zhang:2014rja} 
    Zhang C 2014
    %``Effective field theory approach to top-quark decay at next-to-leading order in QCD,''
    {\it Phys. Rev.} D {\bf 90} 014008
    %[arXiv:1404.1264 [hep-ph]].
    %%CITATION = ARXIV:1404.1264;%%
  %5 citations counted in INSPIRE as of 22 Nov 2014

%\bibitem{Alonso:2013hga} 
%  Alonso R, Jenkins E E, Manohar A V and Trott M 2014
%  %``Renormalization Group Evolution of the Standard Model Dimension Six Operators III: Gauge Coupling Dependence and Phenomenology,''
%  {\it JHEP} {\bf 1404} 159
%  %[arXiv:1312.2014 [hep-ph]].
%  %%CITATION = ARXIV:1312.2014;%%
%  %35 citations counted in INSPIRE as of 22 Nov 2014
\bibitem{Liu:2005dp} 
  Liu J J, Li C S, Yang L L and Jin L G 2005
  %``Next-to-leading order QCD corrections to the direct top quark production via model-independent FCNC couplings at hadron colliders,''
  {\it Phys. Rev.} D {\bf 72} 074018 
  %[hep-ph/0508016].
  %%CITATION = HEP-PH/0508016;%%
  %35 citations counted in INSPIRE as of 23 Nov 2014

\bibitem{Alwall:2014hca} 
  Alwall J, Frederix R, Frixione S, Hirschi V, Maltoni F, Mattelaer O, Shao H -S, Stelzer T, Torrielli P and Zaro M 2014
  %``The automated computation of tree-level and next-to-leading order differential cross sections, and their matching to parton shower simulations,''
  {\it JHEP} {\bf 1407} 079
  %[arXiv:1405.0301 [hep-ph]].
  %%CITATION = ARXIV:1405.0301;%%
  %63 citations counted in INSPIRE as of 05 Sep 2014

\bibitem{Alloul:2013bka} 
  Alloul A, Christensen N D, Degrande C, Duhr C and Fuks B 2014
  %``FeynRules  2.0 - A complete toolbox for tree-level phenomenology,''
  {\it Comput. Phys. Commun.}  {\bf 185} 2250
  %[arXiv:1310.1921 [hep-ph]].
  %%CITATION = ARXIV:1310.1921;%%
  %88 citations counted in INSPIRE as of 14 Sep 2014

\bibitem{Degrande:2014vpa} 
  Degrande C 2014
  Automatic evaluation of UV and R2 terms for beyond the Standard Model Lagrangians: a proof-of-principle
  {\it Preprint 1406.3030 [hep-ph]}
  %%CITATION = ARXIV:1406.3030;%%
  %4 citations counted in INSPIRE as of 14 Sep 2014

\bibitem{aMCfcnc}
  Degrande C, Maltoni F, Wang J and Zhang C 2014
  {\it In preparation}

\bibitem{fcncfit}
  Durieux G, Maltoni F and Zhang C 2014
  {\it In preparation}


\bibitem{AguilarSaavedra:2008zc} 
  Aguilar-Saavedra J A 2009
  %``A Minimal set of top anomalous couplings,''
  {\it Nucl. Phys.} B {\bf 812} 181
  %[arXiv:0811.3842 [hep-ph]].
  %%CITATION = ARXIV:0811.3842;%%
  %164 citations counted in INSPIRE as of 22 Nov 2014

\bibitem{AguilarSaavedra:2009mx} 
  Aguilar-Saavedra J A 2009
  %``A Minimal set of top-Higgs anomalous couplings,''
  {\it Nucl. Phys.} B {\bf 821} 215
  %[arXiv:0904.2387 [hep-ph]].
  %%CITATION = ARXIV:0904.2387;%%
  %73 citations counted in INSPIRE as of 22 Nov 2014

\bibitem{Degrande:2012gr} 
  C.~Degrande, J.~M.~Gerard, C.~Grojean, F.~Maltoni and G.~Servant,
  %``Probing Top-Higgs Non-Standard Interactions at the LHC,''
  JHEP {\bf 1207}, 036 (2012)
  [Erratum-ibid.\  {\bf 1303}, 032 (2013)]
  [arXiv:1205.1065 [hep-ph]].
  %%CITATION = ARXIV:1205.1065;%%
  %35 citations counted in INSPIRE as of 24 Nov 2014


\end{thebibliography}
\end{document}